# Power-Aware Hybrid Intrusion Detection System (PHIDS) using Cellular Automata in Wireless Ad Hoc Networks

**P. KIRAN SREE [1], Dr I Ramesh Babu[2,] Dr J.V.R.Murty[3]
Dr. R.Ramachandran[4], N.S.S.S.N Usha Devi[5]**

1. Associate Professor, Department of Computer Science, S.R.K Institute of Technology, Enikepadu, Vijayawada, India, profkiran@yahoo.com.
2. Professor, Dept of C.S.E, Acharya Nagarjuna University, Guntur.
3. Professor, J.N.T.U, Kakinada.
4. Principal, SVCE, Chennai.
5. Graduate Student of JNTU

**Abstract**

*Ad hoc wireless network with their changing topology and distributed nature are more prone to intruders. The network monitoring functionality should be in operation as long as the network exists with nil constraints. The efficiency of an Intrusion detection system in the case of an ad hoc network is not only determined by its dynamicity in monitoring but also in its flexibility in utilizing the available power in each of its nodes. In this paper we propose a hybrid intrusion detection system, based on a power level metric for potential ad hoc hosts, which is used to determine the duration for which a particular node can support a network-monitoring node. Power –aware hybrid intrusion detection system focuses on the available power level in each of the nodes and determines the network monitors. Power awareness in the network results in maintaining power for network monitoring, with monitors changing often, since it is an iterative power-optimal solution to identify nodes for distributed agent-based intrusion detection. The advantage that this approach entails is the inherent flexibility it provides, by means of considering only fewer nodes for re-establishing network monitors. The detection of intrusions in the network is done with the help of Cellular Automata (CA). The CA's classify a packet routed through the network either as normal or an intrusion. The use of CA's enable in the identification of already occurred intrusions as well as new intrusions.*

## I. Introduction

An intrusion is defined as "any set of actions that attempt to compromise the integrity, confidentiality, or availability of a resource". Intrusions in wireless networks amount to *interception, interruption, or fabrication* of data transmitted across nodes, wherein an intruder node attempts to access unauthorized data.

Intrusion detection is one of key techniques behind protecting a network against intruders. An Intrusion Detection System is a system that tries to detect and alert on attempted intrusions into a system or network, where an intrusion is considered to be any unauthorized or unwanted activity on that system or network [2]. *Ad hoc networks* are particularly prone to such dangers, considering the dynamic and geographically distributed nature of the nodes. Ad hoc networks can hence be classified on the basis of their dynamism as minimally mobile or highly mobile. Thus it requires a combination of both network based intrusion detection and host based intrusion detection systems.

Hybrid intrusion detection systems are inherently reconfigurable, since the agents can easily be migrated to other hosts, and are by themselves lightweight, and thus suit the power sensitive nature of networks such as wireless sensor networks. We adopt a hierarchical model for Intrusion Detection and extend it to include power awareness of individual nodes. We utilize the power level metric PLANE as described in [1], for comparing power levels across nodes for running agent-based network monitoring processes. A complete analysis of possible network threats to general ad-hoc networks is found in [3, 10].We adopt the hierarchical model proposed in [2] and, modify SPAID [1], extending it to provide efficient power aware solution for dynamic networks.

**II. Survey of Related Work**

Numerous detection systems have been proposed to tackle the problem of intrusion in wireless networks some of which are an extension of intrusion detection system in wired networks. Few deal with network based IDS [6] and few with host based IDS [7], all which are based on lightweight agents [5, 6, 7]. Power awareness in mobile ad hoc networks [4] becomes a major issue when considering intrusion detection in larger networks.

**2.1. Triggering Mechanisms**

To protect the network, IDS must generate alarms when it detects intrusive activity on the network. Different IDSs trigger alarms based on different types of network activity.
 The two most common triggering mechanisms are the following:
    1. Anomaly detection
    2. Misuse detection
Discussions regarding the above triggering mechanisms can be found in [3]. Besides implementing a triggering mechanism, the IDS must somehow watch for intrusive activity at specific points within the network. Monitoring intrusive activity normally occurs at the following two locations:

1. Host – Host based IDS
2. Network – Network based IDS

Finally, many intrusion detection systems incorporate multiple features into a single system. These systems are known as hybrid systems. These hybrid intrusion detection systems having their architecture based on agents [2, 9] which travel throughout the network, provide a comprehensive solution.

### 2.2. Host-Based IDS (HIDS)

Host-based systems, the first type of IDS to be developed and implemented, collect and analyze data that originate on a computer that hosts a service, such as a Web server. Once this data is aggregated for a given computer, it can either be analyzed locally or sent to a separate/central analysis machine. One example of a host-based system is programs that operate on a system and receive application or operating system audit logs. These programs are highly effective for detecting insider abuses. Residing on the trusted network systems themselves, they are close to the network's authenticated users. If one of these users attempts unauthorized activity, host-based systems usually detect and collect the most pertinent information in the quickest possible manner. In addition to detecting unauthorized insider activity, host-based systems are also effective at detecting unauthorized file modification. Host-based commercial products include ITA, Squire, and Entercept, to name a few.

### 2.3. Network-Based IDS (NIDS)

Network-based intrusion detection analyzes data packets that travel over the actual network. These packets are examined and sometimes compared with empirical data to verify their nature: malicious or benign. Because they are responsible for monitoring a network, rather than a single host, Network-based intrusion detection systems (NIDS) tend to be more distributed than host-based IDS. Software, or appliance hardware in some cases, resides in one or more systems connected to a network, and is used to analyze data such as network packets. Instead of analyzing information that originates and resides on a computer, network-based IDS uses techniques like "packet-sniffing" to pull data from TCP/IP or other protocol packets traveling along the network. Examples of network-based IDS include Shadow, Snort, Dragon, NFR, RealSecure, and NetProwler.

## 2.4. Hybrid Intrusion Detection System

The two types of intrusion detection systems differ significantly from each other, but complement one another well. The network architecture of host-based is agent-based, which means that a software agent resides on each of the hosts that will be governed by the system. In addition, more efficient host-based intrusion detection systems are capable of monitoring and collecting system audit trails in real time as well as on a scheduled basis, thus distributing both CPU utilization and network overhead and providing for a flexible means of security administration.

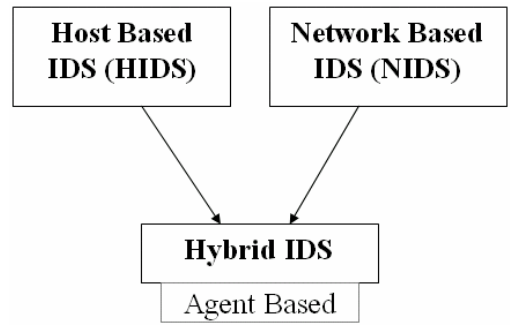

**Fig.1.** Hybrid Intrusion Detection System

In a proper IDS implementation, it would be advantageous to fully integrate the network intrusion detection system, such that it would filter alerts and notifications in an identical manner to the host-based portion of the system, controlled from the same central location (Fig.1). In doing so, this provides a convenient means of managing and reacting to misuse using both types of intrusion detection.

Although network intrusion detection has its merits and certainly must be incorporated into a proper IDS solution, it has historically been incapable of evolving to comply with the growing technology of data communications. Most NIDS perform miserably, if at all, on switched networks, fast networks of speeds over 100 Mbps, and encrypted networks. Furthermore, somewhere in the range of 80 - 85 percent of security incidents originate from within an organization. Consequently, intrusion detection systems should rely predominantly on host-based components, but should always make use of NIDS to complete the defense.
In short, a truly secure environment requires both a network and host-based intrusion detection implementation to provide for a robust system that is the basis for all of the monitoring, response, and detection of computer misuse.

### 2.5. Power-aware Design:

Energy-aware design and evaluation of network protocols requires knowledge of the energy consumption behavior of actual wireless interfaces. But little practical information is available about the energy consumption behavior of well-known wireless network interfaces and device specifications do not provide information in a form that is helpful to protocol developers [1].

Wireless devices must operate for a long period of time, relying on their battery power. Although many developers have looked at extending the life of a wireless system from a hardware point of view, such as directional antennas and improving battery life, power-awareness is a relatively new concept in wireless networking [2]. The metrics that have so far been devised to minimize power can be grouped into two main categories, power-aware and cost-aware metrics. Power-aware metrics aim to minimize the total power needed for intrusion detection when a packet is routed through a network while cost-aware metrics look at methods which extend the nodes' battery lifetime.

Because mobile devices are dependent on battery power, it is important to minimize their energy consumption. The energy consumption of the network interface can be significant, especially for smaller devices. Most research in energy conservation strategies has targeted wireless networks that are structured around base stations and centralized servers, which do not have     the limitations associated with small, portable devices. By contrast, an ad hoc network is a group of mobile, wireless hosts which cooperatively form a network independently of any fixed infrastructure. The multi-hop routing problem in ad hoc networks has been widely studied in terms of bandwidth utilization, but energy consumption has received less attention [12]. So has it in intrusion detection. It is sometimes (incorrectly) assumed that bandwidth utilization and energy consumption are roughly synonymous. Recent studies are that of energy-aware ad hoc routing protocols, particularly for distributed sensor networks. In this context, energy is often treated as an abstract "commodity" for purposes of minimizing cost or maximizing time to network partition.

We believe that energy-aware design and evaluation of network protocols for the ad hoc networking environment requires practical knowledge of the energy consumption behavior of actual wireless devices [4]. In addition, it is important to present this information in a form that is useful to protocol developers: the total energy costs associated with a packet containing some number of bytes of data. Device specifications, which indicate the current draw while transmitting and receiving, are somewhat unhelpful in this respect.

### III. PHIDS & Cellular Automata (CA)

Cellular automata use localized structures to solve problems in an evolutionary way. CA often demonstrates also significant ability toward self-organization that comes mostly from the localized structure on which they operate. By organization, one means that after some time in the evolutionary process, the system exhibits more or less stable localized structures (Wolfram, 2002). This behavior can be found no matter the initial conditions of the automaton.

A CA[4-6], consists of a number of cells organized in the form of a lattice. It evolves in discrete space and time. The next state of a cell depends on its own state and the states of its neighboring cells. In a 3-neighborhood dependency, the next state $q_i(t+1)$ of a cell is assumed to be dependent only on itself and on its two neighbors (left and right) and is denoted as:

$$q_i(t + 1) = f(q_{i-1}(t), q_i(t), q_{i+1}(t)) \quad (1)$$

where, $q_i(t)$ represents the state of the $i^{th}$ cell at $t^{th}$ instant of time, f is the next state function and referred to as the rule of the automata. The decimal equivalent of the next state function, as introduced by Wolfram, is the rule number of the CA cell[8].

**3.1 Fuzzy CA fundamentals:** FCA[2,6] is a linear array of cells which evolves in time. Each cell of the array assumes a state $q_i$, a rational value in the interval [0,1] (fuzzy states) and changes its state according to a local evolution function on its own state and the states of its two neighbors. The degree to which a cell is in fuzzy states 1 and 0 can be calculated with the membership functions. This gives more accuracy in finding the coding regions. In a FCA, the conventional Boolean functions are AND, OR, NOT.

**Dependency matrix for FCA:** Rules defined in Eq. 1 should be represented as a local transition function of FCA cell. That rules (Table 1) are converted into matrix form for easier representation of chromosomes.

**Example 1:** A 4-cell null boundary hybrid FCA with the following rule <238, 254, 238, 252> (that is, <($q_i+q_{i+1}$), ($q_{i-1}+q_i+q_{i+1}$), ($q_i+q_{i+1}$), ($q_{i-1}+q_i$)>) applied from left to right, may be characterized by the following dependency matrix.

While moving from one state to other, the dependency matrix indicates on which neighboring cells the state should depend. So cell 254 depends on its state, left neighbor and right neighbor . Now we represented the transition function in the form of matrix. In the case of complement[5,6,8], FCA we use another vector for representation of chromosome.

Table 1: Rule of FCA

| Non-complemented rules | | Complemented rules | |
| --- | --- | --- | --- |
| Rule | Next state | Rule | Next state |
| 0 | 0 | 255 | 1 |
| 170 | $q_{i+1}$ | 85 | $\overline{q_{i+1}}$ |
| 204 | $q_i$ | 51 | $\overline{q_i}$ |
| 238 | $q_i + q_{i+1}$ | 17 | $\overline{q_i + q_{i+1}}$ |
| 240 | $q_{i-1}$ | 15 | $\overline{q_{i-1}}$ |
| 250 | $q_{i-1} + q_{i+1}$ | 5 | $\overline{q_{i-1} + q_{i+1}}$ |
| 252 | $q_{i-1} + q_i$ | 3 | $\overline{q_{i-1} + q_i}$ |

$$T = \begin{bmatrix} 1 & 1 & 0 & 0 \\ 1 & 1 & 1 & 0 \\ 0 & 0 & 1 & 1 \\ 0 & 0 & 1 & 1 \end{bmatrix}$$

Fig 2: Matrix Representation of Rule

### 3.2 Challenges of any Network Based Intrusion System

The cost of ownership should be lower for an enterprise environment. Network-based IDS must examine all packet headers for signs of malicious and suspicious activity. They have to use live network traffic for real-time attack detection. Therefore, an attacker cannot remove the evidence. They also detect malicious and suspicious attacks as they occur, and so provide faster notification and response. These are not dependent on host operating systems as detection sources. Network-based IDS add valuable data for determining malicious intent.

### 3.3 Genetic Algorithm & CA

The main motivation behind the evolving cellular automata framework is to understand how genetic algorithms evolve cellular automata that perform computational tasks requiring global information processing, since the individual cells in a CA can communicate only locally without the existence of a central control the GA has to evolve CA that exhibit higher-level emergent behavior in order to perform this global information processing .Thus this framework provides an approach to studying how evolution can create dynamical systems in which the interactions of simple components with local information storage and communication give rise to coordinated global information processing.

### 3.4 Crossover and Mutation

Traditional genetic algorithms have been used to identify and converge populations of candidate hypotheses to a single global optimum. For this problem, a set of rules is needed as a basis for the IDS. There is no way to clearly identity whether a network connection is normal or anomalous just using one rule. Multiple rules are needed to identify unrelated anomalies, which means that several good rules are more effective than a single best rule. Another reason for finding multiple rules is that because there are so many network connection possibilities. The *sharing* method reduces the fitness of individuals that have highly similar members and forces individuals to evolve to other local maxima that may be less populated. The similarity metrics used in these techniques can be phenotype to genotype similarity such as Hamming distance between bit representations, or phenotype similarity such as the relation between two network connections in this problem. The latter one is more fitful for finding rules used in IDS. The disadvantage of this approach is that it requires more domain-specific knowledge The mutation operation should be meaningful during evolution. For example, each segment of the IP address should not exceed 255 (decimal representation). Mutations should be done following the requirements specified in Table 1. These limitations can be enforced by defining proper mutation rules.

### 3.5 CA frame work for Intrusion

A computational task T for a CA is now defined as a mapping from initial configurations s0 to sets of answer states Ai,(1<= i <=N) for finite n. An answer state Ai consisting of finite set of m configurations sj (1<= j <=N) which is time invariant which the CA must cycle repeatedly as show in equation 1.

$$T = \sum^{N} \rightarrow \{Ai : 1 < i < N\} \quad - (2)$$

### 3.6 CA Key Distribution

**NIDSWCA** depends on public key servers when using public key cryptography. To simply the design and reduce the load on the endpoints running NIDSWCA, a centralized, trusted Key Distribution Center (KDC) is used in our design. Since NIDSWCA hosts communicate through multicast, potentially every host on the trusted network will get messages addressed to the NIDSWCA group, which it will have to subsequently validate against the sender's public key. This imposes heavy and bursty load on the KDC.

At a theoretical level, cellular automaton Intrusion detection models can be analyzed by much the same methods of statistical mechanics as have been used in trying to derive the Navier-Stokes equations for physical fluids from the microscopic dynamics of real molecules. Fig2 show the Design of CA based Intrusion Detection. The resulting equations for these macroscopic quantities correspond closely with the usual Navier-Stokes equations. Just like a real fluid, however, the cellular automaton model contains definite higher-order corrections, not included in the Navier-Stokes equations. In addition, analytical methods provide only approximate values for parameters such as viscosity; accurate values must be obtained from explicit computer simulations.

### 3.7  Basic Algorithm

CA Tree Building (Assuming K CA Basins)

Input:    Intrusion parameters (constraints)
Output: CA Based  inverted tree
Step 0: Start.
Step 1: Generate a CA with *k* number of CA basins
Step 2: Distribute the parameters into k CA basins
Step 3: Evaluate the distribution in each closet basin
Step 4: Calculate the Rq( Malicious Index)
Step 5: Swap the more appropriate features to the bottom leaves of the inverted tree.
Step 6: Stop.

### 3.8 CA Agent System

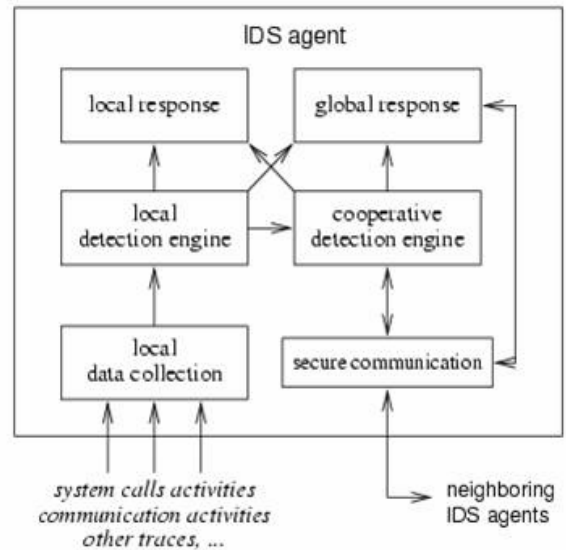

Fig 3 : IDS Agent System

The main feature which is achieved when developing CA Agent systems(fig 3), if they work, is flexibility, since a CA Agent system can be added to, modified and reconstructed, without the need for detailed rewriting of the application. These systems also tend to be rapidly self-recovering and failure proof, usually due to the heavy redundancy of components and the self managed.

The agent-based model proposed in [10] approaches the IDS problem with a technique that handles intrusions with an agent running on each system. Further, the model in [10] is not suitable for a power-aware IDS , since such a system warrants energy consumption in systems irrespective of their current battery levels, i.e. it suggests an IDS without considering the feasibility of the assumption that network monitoring and analysis is justified in nodes with minimal power, such as robust wireless sensor networks (WSN).

### 3.9 Modular IDS Architecture

The IDS we consider is built on a mobile agent framework as in [1]. It is a non-monolithic system and employs several sensor agents that perform certain functions, such as:

• *Network monitoring*: Only certain nodes will have sensor agents for network packet monitoring, since we are interested in preserving the total computational power and battery power of mobile hosts.

• *Host monitoring:* Every node on the mobile ad hoc network will be monitored internally by a host-monitoring agent. This includes monitoring system-level and application-level activities.

• *Decision-making:* Every node will decide the intrusion threat level on a host-level basis. Certain nodes will collect intrusion information and make collective decisions about network level intrusions.

• *Action:* Every node will have an action module that is responsible for resolving intrusion situation on a host (such as locking out a node, killing a process, etc).

A hierarchy of agents has been devised in order to achieve the above goals. We will adapt the hierarchy for our purposes. There are three major agent classes as used in [2], categorized as monitoring, decision-making and action agents. Some are present on all mobile hosts, while others are distributed to only a select group of nodes, as discussed further. The monitoring agent class consists of packet, user, and system monitoring agents. The following diagram shows the hierarchy of agent classes.

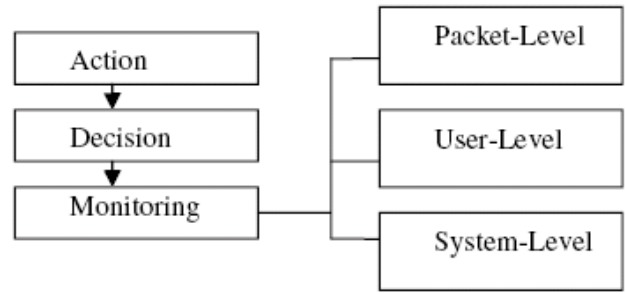

**Fig.4.**Typical agent hierarchy, depicting the multi-level decision making process for intrusion detection

A detailed discussion of each of the modules and its functionalities can be found in [2].

**3.10 Agent Distribution**

As a modification over the previous approach for agent distribution [1], the nodes on a wireless ad-hoc network, that are elected as network monitors will include the action and decision making modules. To save resources, some of the functionality must be distributed efficiently to a (small) number of nodes. The decision making module incorporates the energy metric Power Loss/Availability for Network-monitoring Estimate (PLANE), a node-specific measure of the mean

P. KIRAN SREE,I Ramesh Babu,J.V.R.Murty
R.Ramachandran, N.S.S.S.N Usha Devi

power loss per node for running the network monitoring agent. PLANE can directly be related to the wireless protocol used, mean number of wireless links for the specific node, average node maintenance energy consumption, and the battery power remaining. PLANE ultimately estimates the duration the node can last on the same power without replenishment. To calculate the power consumption metrics such as those in are often used. The reception costs are multiplied by the number of links for the node to yield an average reception cost, to which we add the average cost of sending a message. Thus, these costs are dependent on the density of the network and the routing/link exchange protocols used.

### 3.11 Calculating PLANE

The calculation of PLANE involves calculating the duration for which the node can continue to support a network monitor along with its normal operations. We therefore calculate PLANE by calculating the time for which node can last as the network monitoring node as shown below in Equation 1.

$$PLANE = \frac{BPR}{TEC_{nm}} \qquad (1)$$

In Equation 1, BPR is the total Battery Power Remaining at the instant of node selection and TECnm is the Total Energy Consumption with network monitoring node processes running. In the absence of measurement of exact networking monitoring energy consumption, we assume PLANE as PLANE'. The value PLANE' is typically available directly from most distributed wireless networks, such as sensor networks, and hence finds a presence in the above calculation.

$$PLANE' = \frac{BPR}{TEC} \qquad (2)$$

TEC is the total energy consumption before the node is selected for network monitoring. PLANE can be tailored to suit the needs of the type of network monitoring required and the nature of the actual node on which it runs. TEC values are represented by different wireless nodes running in different ad hoc modes which consume between 741 mW and 843 mW.

### 3.12 The PHIDS Algorithm

In PHIDS, we deal with multi-hop network monitoring clustered node selection, similar to SPAID. This type of a node selection has its inherent advantages in allowing complete coverage of all nodes and links in a network, but with an added factor of redundancy in the collection of intrusion detection data. The algorithm presented here is a power efficient variation over the previous SPAID, where the increment in hop radius and re-running of SPAID was considered for the whole topology, after a particular drop in power for certain monitoring nodes.

This approach considers each of the initially allocated monitors and the nodes they monitor to be a single tree, with the monitoring node as a root and the nodes being monitored as its child. The root along with its child nodes contribute to individual clusters. Thus the whole large topology gets divided to tree structured clusters only for intrusion detection purpose. Thus once a node is selected and allotted as a monitor initially for a set of nodes, they form an individual cluster.

After such clusters are established, when any drain in power levels take place to the monitors, any other child node with higher battery power for monitoring can take charge of that cluster. Since only certain clusters are active for certain period of time, it is enough if their root nodes i.e., their monitoring nodes get re-arranged (within the cluster), instead of running the whole node selection process as in the case of SPAID which is inefficient.

The node selection process should be done for a network as a whole only when no single node within a cluster is competent enough to monitor or when a new node with higher PLANE value enters the existing network.

The PHIDS algorithm uses the agent hierarchy presented in Fig.2, with a significantly adapted node selection mechanism to incorporate power-awareness, and is best detailed by the following eight steps.

*Step 1: Set PLANE threshold*. Set a constraint on the PLANE value of nodes which are allowed to compete for becoming a network monitoring node.

*Step 2: PLANE Calculation and PLANE Ordered List (POL).* Arrange the different nodes in increasing values of PLANE as calculated previously, for all nodes which satisfy the PLANE Constraint.

*Step 3: Hop Radius.* Set the hop radius to one initially, and increment for each insufficient node selection with the current hop radius.

*Step 4: Expand Working Set of Nodes.* Consider node selection incrementally, initially from the first node, (node with highest PLANE), to finally the set of all nodes in the network, incrementing the set of nodes under consideration by one node each time. We call this set the working set (WS) of nodes. The WS is expanded only if the addition leads to an increase in number of represented nodes.

*Step 5: Voting. The voting scheme* for Node Selection, is similar to that in [2], except that we limit the candidates to just the nodes which are part of WS.
*Step 6: Check acceptability of nodes.* If all links/nodes are not represented by the set of nodes covered by the voting scheme, then we expand the WS and repeat the process from Step 4. If WS equals the POL, then increment the hop radius, and repeat from Step 3. It is suggested that the increment in hop radius be considered a final resort, as it effectively increases the amount of processing per monitoring node.
*Step 7: Cluster Setup.* Set individual clusters with the nodes in the working set as root and the nodes being monitored by it as child nodes.
*Step 8: Re-run.* Changes in power levels of the root nodes in each cluster will be signaled to the child and the vote count as in step 5 takes place within the cluster to form a new monitoring node.

Steps 1 to 6 are similar to that of SPAID but are suitable even to highly mobile networks, since decision making module does not rest with very few nodes and only clusters are altered on each addition of a new node.

### 3.13 An Example

Consider an ad hoc network with 9 nodes and their PLANE values as given below.

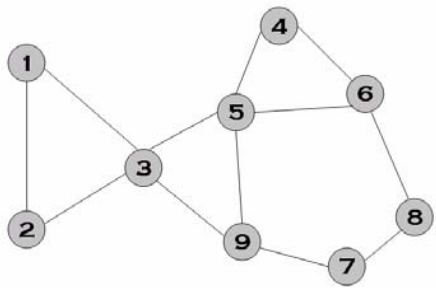

Node 1 : 5.5    Node 6 : 4.1
Node 2 : 7.2    Node 7 : 7.5
Node 3 : 9.0    Node 8 : 5.7
Node 4 : 8.5    Node 9 : 7.0
Node 5 : 5.0

**Fig.5.** An example network

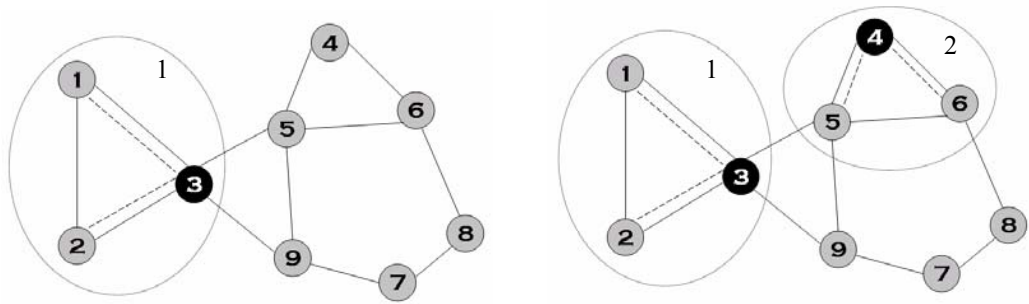

**Fig.6.** Example 1 with WS = {3}    **Fig.7.** Example 1 with WS = {3, 4}

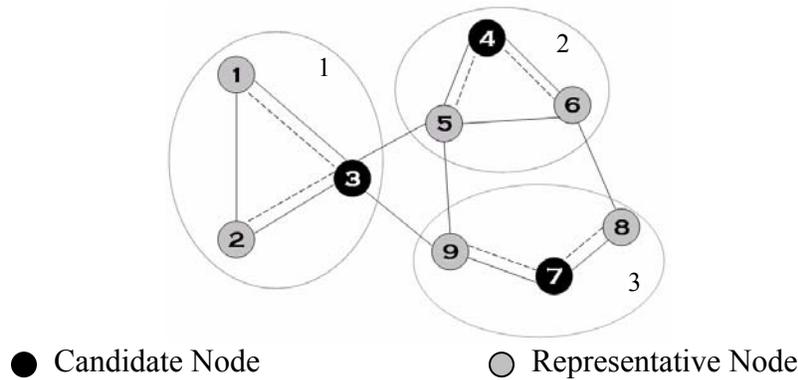

● Candidate Node         ○ Representative Node

**Fig.8.** Example 1 with WS = {3, 4, 7} Final Selection

As shown from Fig 3 to Fig 6, the node selection algorithm mentioned above considers the power availability of each of the nodes in the network. The working set {3, 4, 7} are the monitors and they form the cluster heads. The dotted lines represent the monitoring function that they perform using agents. Take for example Fig.4., node 3 is the monitor, detecting intrusions over 1 and 2, thus forming a cluster represented by the circle.

   Step 8 of the algorithm proves to provide an energy efficient solution whenever one or two clusters actively participate in transmitting, receiving and routing. Variation in power levels take place much more in those nodes than those of any other clusters. It is sufficient enough to change the monitoring node only in those clusters instead of troubling the whole setup. Supposing cluster 1 with root 3 is constantly participating in routing, the power level will significantly decrease in 3, since it has to monitor the packets transmitted across it and its child for intrusions. At one point of time the BPR of node 3 may drop below that of node 2. Thus the rearrangement within the cluster takes place, with node 2 becoming the root i.e. the monitor and node 3 and 1 become its children, as shown below.

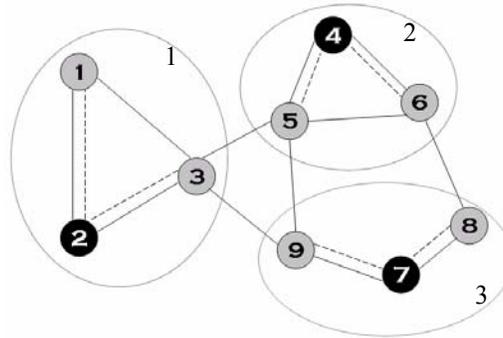

**Fig.9.** Example 1 with WS = {3, 4, 7} with Cluster 1 Re-established

On constant participation in the network activities, if all the nodes in a particular cluster, lose their power below the threshold, then a re-run of the entire node selection algorithm can be done. Since a new node can enter anytime into the existing network, the node selection algorithm should be employed to find out the network monitors.

**IV. Intrusion Detection Module using CA:**

Cellular Automata s (CA's) represent a generalized linear classifier that looks for the maximum margin hyper plane between two classes in the feature space i.e. the CA's try to find a hyper plane between two classes which maximizes the minimum distance between the hyper plane and the data points.

**Maximum Margin Hyperplane**

The optimal position of the class boundary is obtained as a linear combination of some training samples that are placed near the boundary itself, and are called Cellular Automata s.

P. KIRAN SREE,I Ramesh Babu,J.V.R.Murty
R.Ramachandran, N.S.S.S.N Usha Devi

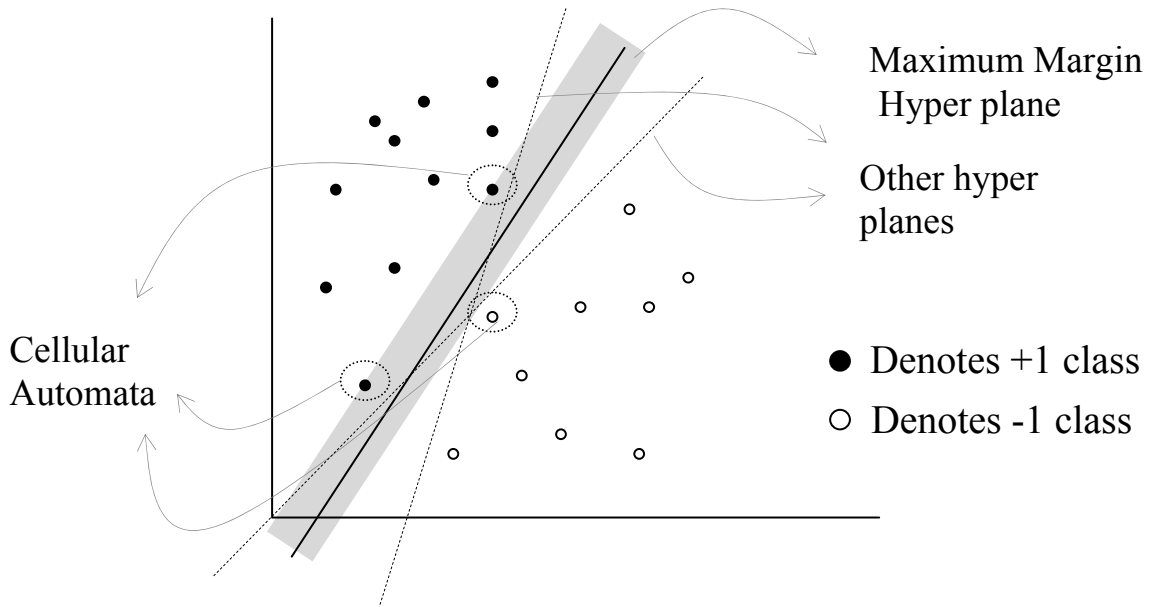

**Fig 10:** CA as a Linear Classifier

The hyper planes are defined by the linear set of equations given by:

$$w \cdot x + b = 0, \quad w \in \mathbb{R}^d, \quad b \in \mathbb{R}$$

Each input x is subject to the decision function O(x):

O(x) = sign (w.x +b)

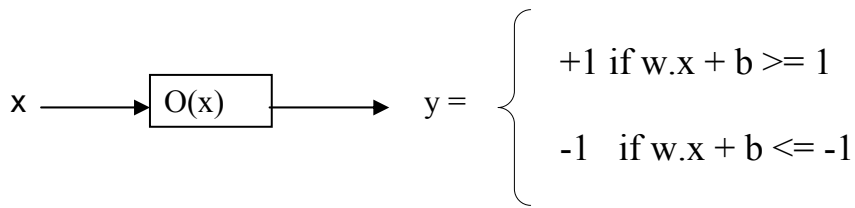

The margin width of the hyper plane is computed by considering the plus plane and minus plane

Plus plane = {x: w.x +b = +1}
Minus plane = {x: w.x +b = -1}

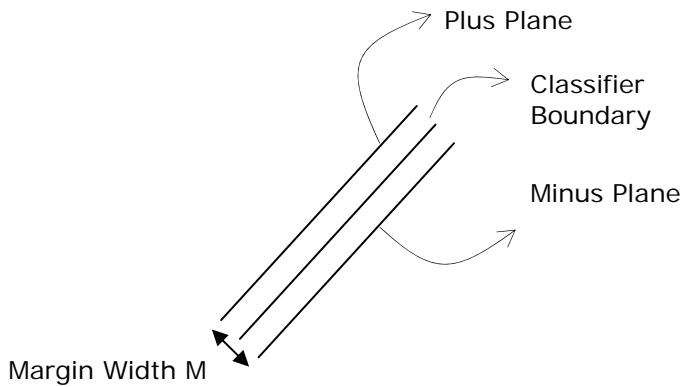

**Fig 11:** Margin Width of the CA Classifier

The perpendicular distance of each of the plus and the minus planes respectively from the classifier boundary are given $\frac{|1-b|}{||w||}$ and $\frac{|-1-b|}{||w||}$. Hence the margin width is $\frac{|1-b+1+b|}{||w||} = \frac{2}{||w||}$ and the pair of hyper planes that gives the maximum margin can be found by minimizing w2..This is a quadratic programming (QP) problem. It is solved by LaGrange Formulation of the margin width equation. We have seen an overview of how to use CA for binary classification purpose.

P. KIRAN SREE,I Ramesh Babu,J.V.R.Murty
R.Ramachandran, N.S.S.S.N Usha Devi

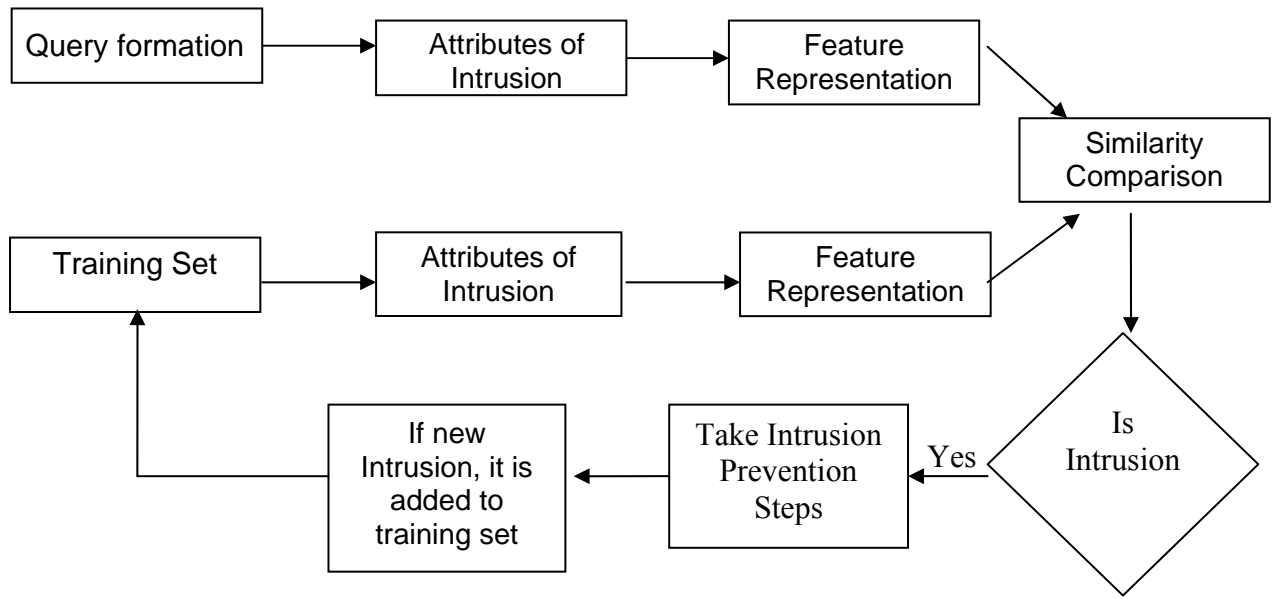

**Fig 12.** Flow of Controls in Intrusion Detection using Cellular Automata s

## DETAILS OF ALGORITHM

There are two phases to the implementation of classification problem using CA's. They are the

- Training or Learning phase
- Testing or Recognition phase

A detailed description of the algorithm is shown in the figure below. It has the two phases-training phase and testing.

**Training or Learning phase**

In the training phase the optimal hyper planes for each of the binary classifiers are constructed based on the training set data. The system is trained with a lot of sample intrusions and their attributes. These images form the basis in identifying the image the user queries. The attribute vector for each of these intrusions is found and is stored in a feature database.

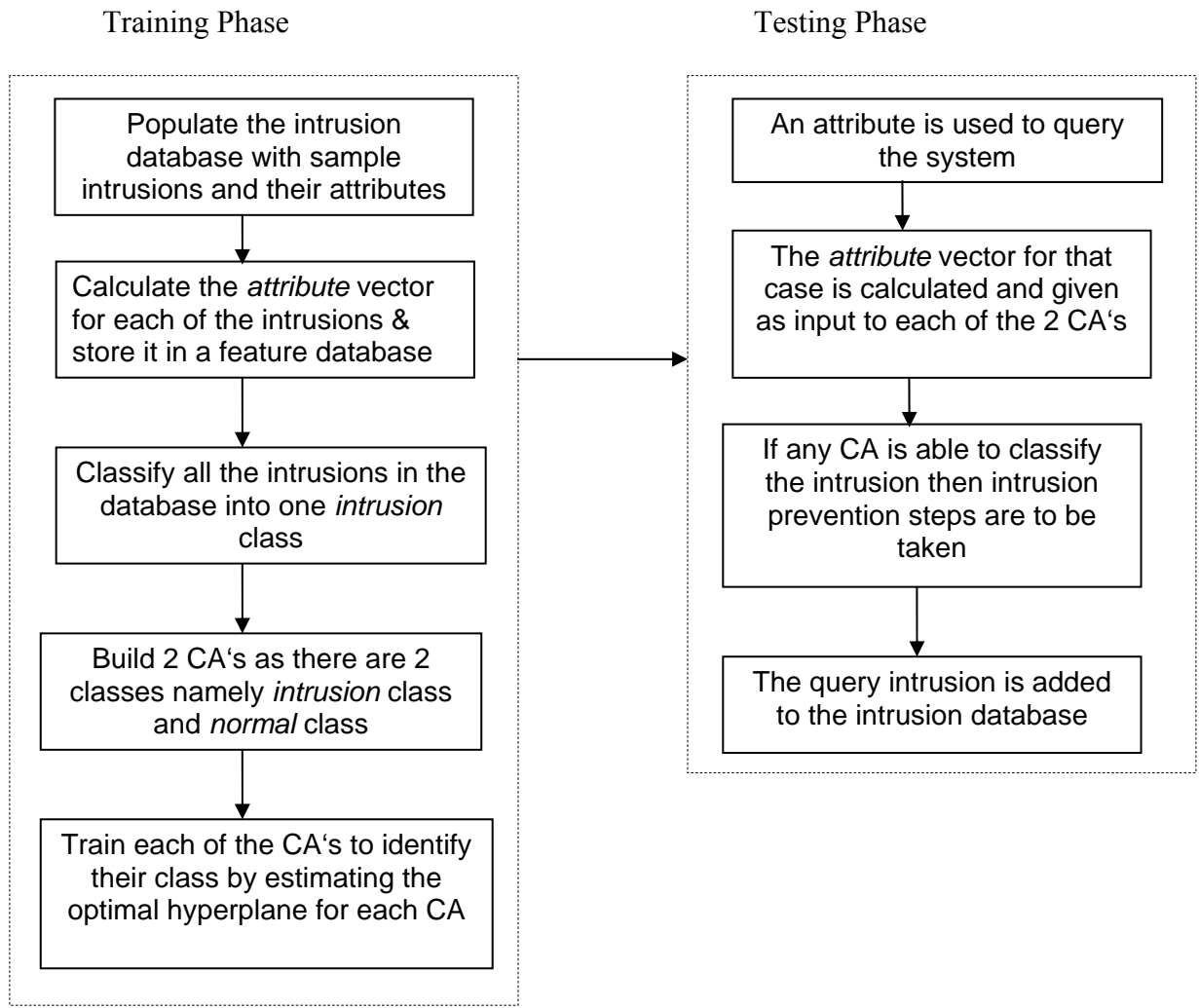

The instances belonging to a similar class are grouped into a single category. There are 2 classes namely INTRUSION class and NORMAL class, therefore build 2 CA's. Each CA is trained to identify its class by estimating its optimal hyperplane for each CA.

 **Testing or Recognition phase**

In the testing phase the attributes of the instance are used to query. The attribute vector is calculated for the image and is given as input to the pool of trained CA's which identifies the class to which the instance belongs and takes the necessary steps.

## V. Experimental Results & Performance Comparison:

Evaluating the extended algorithm, PHIDS, in terms of power, results in much better utilization of the available power.

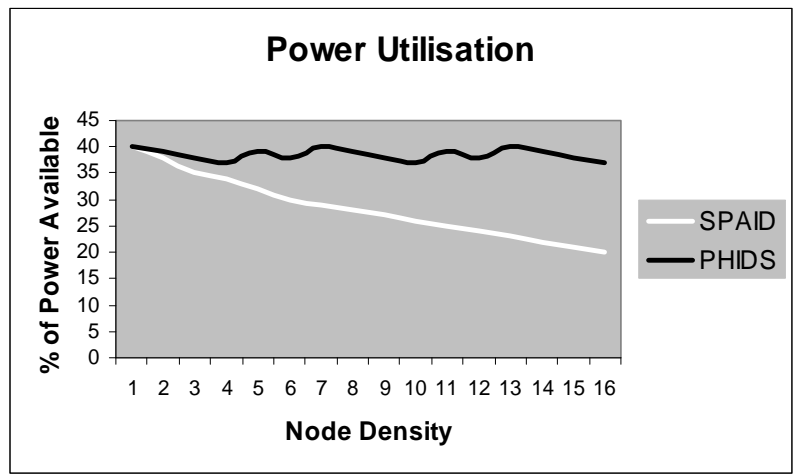

**Fig.13.** Performance Comparison – SPAID vs. PHIDS

Splitting up of larger networks to clusters and manipulating power levels and thresholds within them provides a power optimal solution than that of SPAID, which requires the entire network. Also SPAID [1] was considered only for minimally mobile networks with increments in hop radius for more dense networks, while PHIDS with tree based clusters can prove efficient even in the case of dynamic networks.

Each data point is described by 41 features. Note that some attributes are continuous and some are nominal. Since the clustering and classification algorithms require continuous values, these nominal values will be first converted to continuous. Table 2,3,4 shows that NIDSWCA consumes less time for training .

Table 2: Results for Un Supervised CA with network 1

| Algorithm | Adaptive Learning Rate | Resilient Back propagating | Unsupervised CA |
|---|---|---|---|
| Train Accuracy | 40.3% | 76% | 89% |
| Test Accuracy | 24% | 56% | 80% |
| Performance | 0.33 | 0.004 | 0.5 |
| Time to train | 60 mins | 65 mins | 50 mins |

Table 3: Results for UN Supervised CA with network 2

| Algorithm | Adaptive Learning Rate | Resilient Back propagating | Unsupervised CA |
|---|---|---|---|
| Train Accuracy | 40.3% | 76% | 89% |
| Test Accuracy | 24% | 56% | 80% |
| Performance | 0.33 | 0.004 | 0.5 |
| Time to train | 60 mins | 65 mins | 50 mins |

Table 4 Results for Un Supervised CA with network

| Algorithm | Adaptive Learning Rate | Resilient Back propagating | Unsupervised CA |
|---|---|---|---|
| Train Accuracy | 40.3% | 76% | 89% |
| Test Accuracy | 24% | 56% | 80% |
| Performance | 0.33 | 0.004 | 0.5 |
| Time to train | 60 mins | 65 mins | 50 mins |

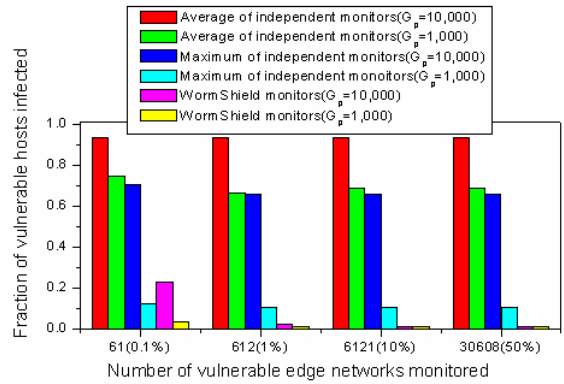

Fig 14: Vulnerable Edges VS Host Infected

Fig 4 shows the graphs which takes the parameter G s( Average number of independent monitors) and depicts the vulnerable hosts percentage of effecting. A CA IDS monitors host and server event/sys logs from multiple sources for suspicious activity. Host IDS are best placed to detect computer misuse from trusted insiders and those who have already infiltrated your network. Okay, IMHO what I have just described is an event log manager, a true Host IDS will apply some signature analysis across multiple events/logs and/or time, heuristically profiling is another useful way to spot nefarious activity.

The algorithm was trained with the first data set. Then we tested the trained algorithm (4. 3) with the same data set and got the following results. We have developed three classifier for dealing the problem of intrusion. Fig 5, 6, 7 depicts the findings.

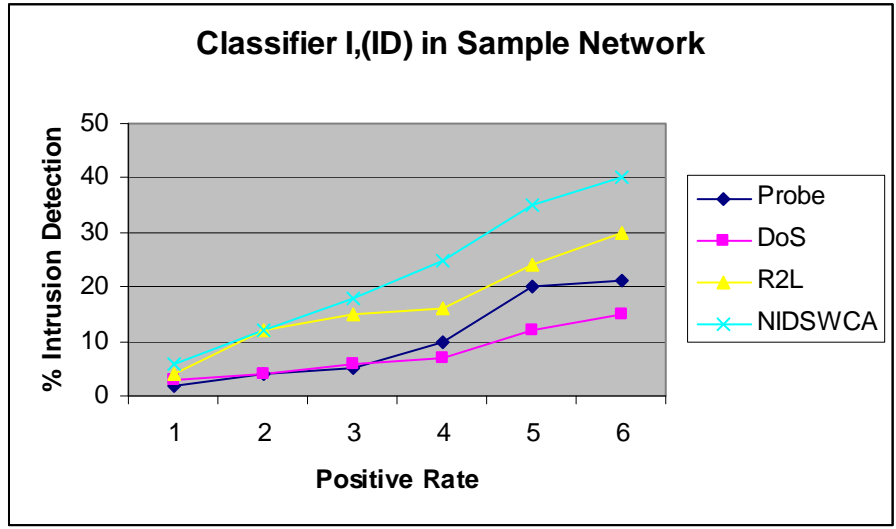

Fig 15: Intrusion Detection Vs Positive Rate in   Sample Network

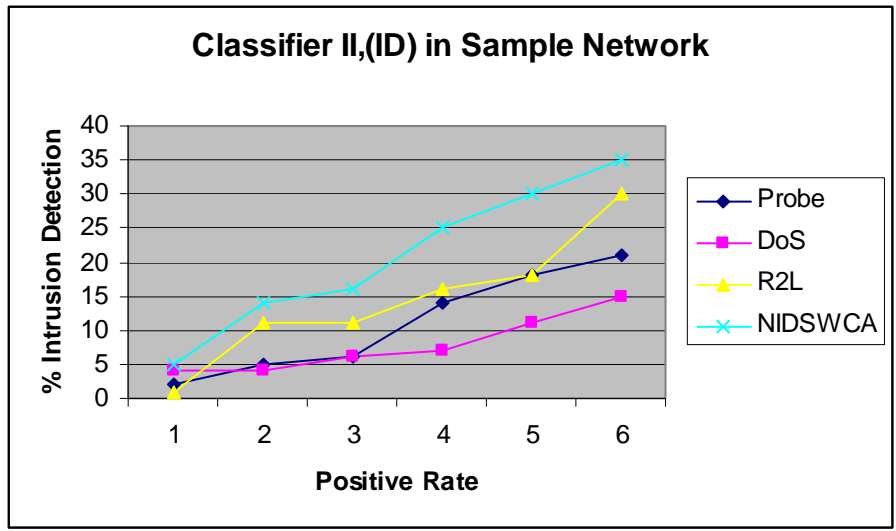

Fig1 6: Intrusion Detection Vs Positive Rate in   Sample Network

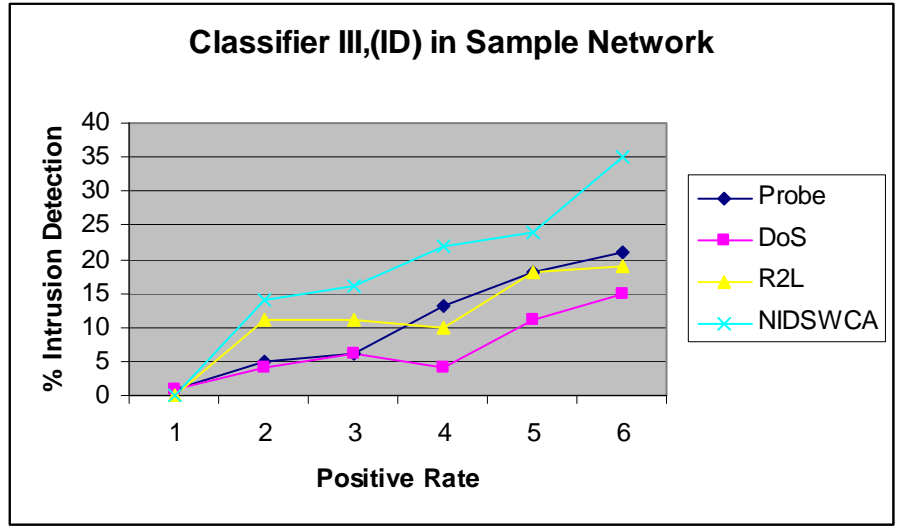

Fig 17: Intrusion    Detection Vs Positive Rate in   Sample Network

## VI. Concluding Remarks

In this paper, sufficient base has been provided to understand the efficient functioning of the PHIDS Algorithm in determining the duration for which a node can support a network monitoring function in wireless ad hoc networks. The preliminary results show that the PHIDS Algorithm gives good results on sparse as well as dense mobile networks. As it is evident from the power utilization performance evaluation the PHIDS algorithm proves to be scalable and even more efficient as the number of nodes increases i.e. as the size of the wireless ad hoc network increases. Re-run over the entire network for node selection needs to be done often only with changing network topologies.

## VII. References

1. 'A System for Power-Aware Agent-Based Intrusion Detection (SPAID) in Wireless Ad Hoc Networks', T. Srinivasan, Jayesh Seshadri, J.B. Siddharth Jonathan,and Arvind Chandrasekhar, Springer-Verlag Berlin Heidelberg 2005.
2. Kachirski, O. and Guha, R.: Efficient Intrusion Detection using Multiple Sensors in Wireless Ad Hoc Networks", 36th Annual Hawaii International

Conference on System Sciences (HICSS'03) - Track 2 , January 06 - 09, 2003.
3. *Zhou, L., and Haas, Z.J.: Securing Ad Hoc Networks, IEEE Networks Special Issue onNetwork Security. November,1999.*
4. *Ahmed Safwat et al, Handbook of Ad hoc Wireless Networks, CRC Press, Dec. 2002,* 'Power-Aware Wireless Mobile Ad hoc Networks', Ahmed M. Safwat, Hossam S. Hassanein, and Hussein T. Mouftah
5. D. Dasgupta and H. Brian, "Mobile Security Agents for Network Traffic Analysis", Proceedings of DARPA Information Survivability Conference & Exposition II, 2001. DISCEX '01, Volume: 2, 2001, pp. 332–340.
6. G. Helmer, J. Wong, V. Honavar, L, Miller, "Lightweight Agents for Intrusion Detection", Technical Report, Dept. of Computer Science, Iowa State University, 2000.
7. M.C. Bernardes and E. Santos Moreira, "Implementation of an Intrusion Detection System based on Mobile Agents", Proceedings of International Symposium on Software Engineering for Parallel and Distributed Systems, 2000, pp. 158-164.
8. Tao, J., Ji-ren, L., and Yang, Q.: "The Research on Dynamic Self-Adaptive Network Security Model Based on Mobile Agent", Proceedings of 36th International Conference on Technology of Object-Oriented Languages and Systems, 2000.
9. Bernardes, M.C., and Moreira, E.S.: "Implementation of an Intrusion Detection System based on Mobile Agents", Proceedings of International Symposium on Software Engineering for Parallel and Distributed Systems, pp. 158-164, 2000.
10. Zhang, Y., and Lee, W.: "Intrusion Detection in Wireless Ad-Hoc Networks", Proceedings of the 6th Annual International Conference on Mobile Computing and Networking, MobiCom, pp. 275-283, 2000.
11. Feeney L.M., and Nilsson, M.: Investigating the Energy Consumption of a Wireless Network Interface in an Ad Hoc Networking Environment, Proceedings of IEEE INFOCOM, 2001.
12. Power Control in Ad-Hoc Networks: Theory, Architecture, Algorithm and Implementation of the COMPOWProtocol , Swetha Narayanaswamy, Vikas Kawadia, R. S. Sreenivas and P. R. Kumar.

13. SVM-based Intrusion Detection System for Wireless Ad Hoc Networks, Hongmei Deng, Qing-An Zeng, and Dharma P. Agrawal

14. Intrusion Detection Systems Using Decision Trees and Cellular Automata Vector Machines,Sandhya Peddabachigari, Ajith Abraham, Johnson Thomas

15. AAnderson, D., Frivold, T., Valdes, A.: Next neration intrusion detection expert system (NIDES): a summary. Technical Report SRI-CSL-95-07. Computer Science Laboratory, SRI International, Menlo Park, CA (May 1995)

16. Axelsson, S.: Research in intrusion detection systems: a survey. Technical Report TR 98-17 (revised in 1999). Chalmers University of Technology, Goteborg, Sweden (1999)

17. H. Abelson, D. Allen, D. Coore, C. Hanson, G. Homsy, T. Knight, R. Nagpal, E. Rauch, G. Sussmanand R. Weiss, Amorphous Computing, Communications of the ACM, Volume 43, Number 5, p. 74-83. May 2000.

18. Qing Cao, Tarek Abdelzaher, Tian He, and John Stankovic. Towards optimal sleep scheduling in sensor networks for rare-event detection. In Proceedings, IPSN '05, page 4, Piscataway, NJ, USA, 2005. IEEE Press.

19. Chao Gui and Prasant Mohapatra. Power conservation and quality of surveillance in target tracking sensor networks. In Proceedings, MobiCom '04, pages 129.143, New York, NY, USA, 2004. ACM Press.

20. P.Kiran See, al. "Improving Quality of Clustering using Cellular Automata for Information retrieval"  Journal of Computer Science 4 (2): 167-171, 2008.